\documentclass[12pt,preprint]{aastex}
\usepackage{psfig} 

\shorttitle{A Survey of 56 Mid-latitude EGRET Error Boxes for Radio
Pulsars} 
\shortauthors{Crawford et al.}

\begin{document}

\title{A Survey of 56 Mid-latitude EGRET Error Boxes for Radio
Pulsars}

\author{Fronefield Crawford\altaffilmark{1,2}, 
Mallory S. E. Roberts\altaffilmark{3,4},
Jason W. T. Hessels\altaffilmark{3}, 
Scott M. Ransom\altaffilmark{3,5}, 
Margaret Livingstone\altaffilmark{3}, 
Cindy R. Tam\altaffilmark{3},
Victoria M. Kaspi\altaffilmark{3}}

\altaffiltext{1}{Department of Physics and Astronomy, Franklin \&
Marshall College, Lancaster, PA 17604, USA; email: fcrawfor@fandm.edu}

\altaffiltext{2}{Department of Physics, Haverford College, Haverford,
PA 19041, USA}

\altaffiltext{3}{Department of Physics, McGill University, Montreal,
QC H3A 2T8, Canada}

\altaffiltext{4}{Eureka Scientific, Inc., 2452 Delmer Street, Suite
100, Oakland, CA 94602, USA}

\altaffiltext{5}{National Radio Astronomy Observatory, 520 Edgemont
Rd., Charlottesville, VA 22903, USA}

\begin{abstract}
We have conducted a radio pulsar survey of 56 unidentified
$\gamma$-ray sources from the 3rd $EGRET$ catalog which are at
intermediate Galactic latitudes ($5^{\circ} < |b| < 73^{\circ}$). For
each source, four interleaved 35-minute pointings were made with the
13-beam, 1400-MHz multibeam receiver on the Parkes 64-m radio
telescope.  This covered the 95\% error box of each source at a
limiting sensitivity of $\sim 0.2$~mJy to pulsed radio emission for
periods $P\ga 10$~ms and dispersion measures $\la 50$ pc cm$^{-3}$.
Roughly half of the unidentified $\gamma$-ray sources at $|b| >
5^{\circ}$ with no proposed active galactic nucleus counterpart were
covered in this survey.  We detected nine isolated pulsars and four
recycled binary pulsars, with three from each class being new.  Timing
observations suggest that only one of the pulsars has a spin-down
luminosity which is even marginally consistent with the inferred
luminosity of its coincident $EGRET$ source.  Our results suggest that
population models, which include the Gould belt as a component,
overestimate the number of isolated pulsars among the mid-latitude
Galactic $\gamma$-ray sources and that it is unlikely that Gould belt
pulsars make up the majority of these sources. However, the
possibility of steep pulsar radio spectra and the confusion of
terrestrial radio interference with long-period pulsars ($P \ga
200$~ms) having very low dispersion measures ($\la 10$ pc cm$^{-3}$,
expected for sources at a distance of less than about 1~kpc) prevent
us from strongly ruling out this hypothesis.  Our results also do not
support the hypothesis that millisecond pulsars make up the majority
of these sources. Non-pulsar source classes should therefore be
further investigated as possible counterparts to the unidentified
$EGRET$ sources at intermediate Galactic latitudes.
\end{abstract}


\section{Introduction}

Determining the nature of Galactic $\gamma$-ray sources with energies
above 100 MeV is one of the outstanding problems in high-energy
astrophysics. The $EGRET$ telescope on the Compton Gamma-Ray
Observatory, which was active from 1991 to 1999, identified about half
a dozen of the brightest $\gamma$-ray sources in the Galactic plane as
young pulsars \citep{tbb+99}.  It also demonstrated that most of the
sources at low Galactic latitudes ($|b| \la 5^{\circ}$) are associated
with star forming regions, and hence may be pulsars, pulsar wind
nebulae, supernova remnants, winds from massive stars, or high-mass
X-ray binaries \citep{kc96,yr97,rbt99}. In addition, molecular clouds
can either be sources of $\gamma$-rays or enhance the production of
$\gamma$-rays by particles produced by the source classes mentioned
above \citep{aha01}.  Various targeted multi-wavelength campaigns to
identify low-latitude sources have discovered a number of likely
counterparts \citep{rrk01,hcg+01, rhr+02,brrk02,hgc+04}. The recent
Parkes Multibeam Survey has also discovered several new pulsars
coincident with $EGRET$ $\gamma$-ray sources; these pulsars have spin
characteristics that are similar to those of the known $\gamma$-ray
pulsars \citep{dkm+01,kbm+03}.

While there are many candidate counterparts to $EGRET$ sources at low
latitudes, there are few firm identifications owing to the large
positional uncertainties of the sources (typically $\sim 1^{\circ}$
across). In general, a timing signature, such as a pulse detection, is
necessary to be certain of a source identity.  Since young pulsars
tend to be noisy rotators, extrapolating a pulse ephemeris reliably
back to the era of the $EGRET$ observation is generally not possible.
With the improved resolution and sensitivity of the upcoming $AGILE$
and $GLAST$ missions, the low-latitude $EGRET$ sources should be more
easily identified.

There are estimated to be between 50 and 100 sources detected by
$EGRET$ at mid-Galactic latitudes which are associated with our
Galaxy.  As a class, these sources tend to be fainter and have steeper
spectra than those at low latitudes \citep{hbb+99}.  Their positional
uncertainty is therefore on average even greater ($\sim 1.5^{\circ}$
across) than it is for the low-latitude sources. These mid-latitude
sources have a spatial distribution which is similar to the Gould belt
of local regions of recent star formation plus a Galactic Halo
component \citep{gre00,gre01}.  The Gould belt provides a natural
birth place for many nearby ($\la 0.5$~kpc), middle-aged pulsars
similar to Geminga \citep{hh92}. Both the outer gap \citep{yr95} and
polar cap \citep{hz01} models of pulsar emission suggest that many of
these pulsars should be detectable in $\gamma$-rays but that the
majority should have their radio beams missing Earth. However, if
predictions from recent models are realistic, then between 25\% and
50\% of $\gamma$-ray pulsars might still be visible to us as radio
pulsars \citep{gvh04,czlj04}.

The mid-latitude $EGRET$ source distribution is also similar to the
distribution of recycled pulsars in the Galactic field
\citep{rom01}. The fastest millisecond pulsars (MSPs) can have
spin-down luminosities ($\dot{E} \propto \dot{P}/P^{3}$) and
magnetospheric potentials similar to those of young pulsars. There has
been one possible detection of $\gamma$-ray pulsations from an MSP
\citep{khv+00} and some preliminary modeling of that emission
\citep{hum05}. If a significant fraction of the mid-latitude sources
are MSPs at typical Galactic distances, many should be detectable as
radio pulsars \citep{sgh05}. Since MSPs tend to be in binary systems,
$GLAST$ will not be sensitive to them in blind searches (owing to
computational reasons associated with the very long integration times
and the large number of trials required to search the parameter
space).

Here we describe a radio pulsar survey of 56 unidentified sources from
the 3rd $EGRET$ catalog (3EG) \citep{hbb+99} which are at intermediate
Galactic latitudes ($5^{\circ} < |b| < 73^{\circ}$).  The survey used
the 1400-MHz, 13-beam multibeam receiver \citep{swb+96} on the 64-m
radio telescope in Parkes, Australia to search for pulsed
emission. This receiver has been used very successfully to find
pulsars in a number of recent radio pulsar surveys \citep{mlc+01,
ebs+01, kbm+03, mfl+06, bjd+06}.  Discovery of radio pulsar
counterparts to these $EGRET$ sources would not only provide
interesting systems for individual study and establish the
identifications of the target sources (e.g., Roberts et
al. 2002\nocite{rhr+02}), but it would also help resolve outstanding
questions about the pulsar emission mechanism and the physical origin
of pulsar radiation at different wavelengths (see, e.g., Harding et
al. 2004 and references therein\nocite{hgg+04}).

\section{Survey Parameters and Data Processing}

\begin{figure} 
\centerline{\psfig{figure=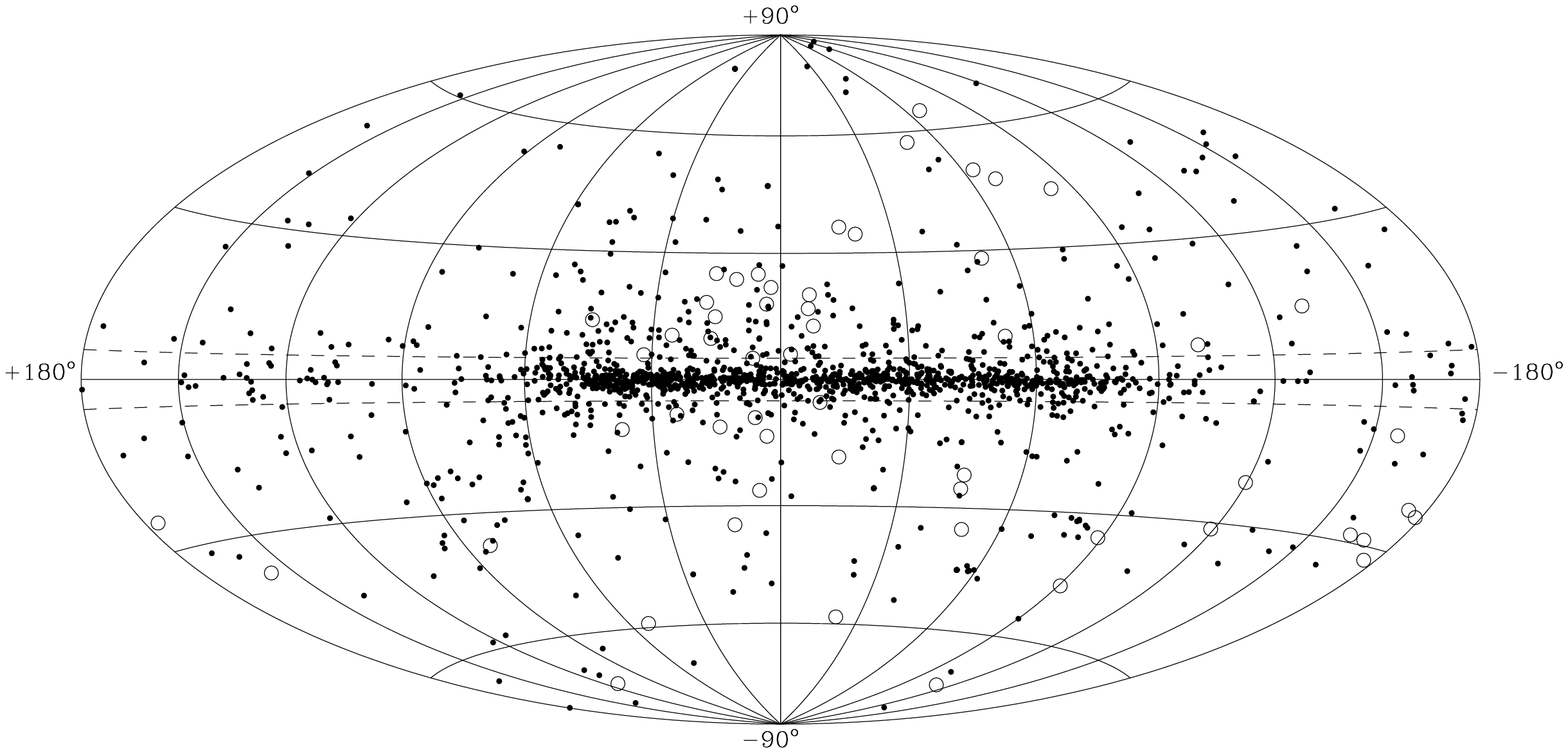,angle=90,width=7in}}
\caption{Aitoff plot in Galactic coordinates of the locations of the
56 unidentified $EGRET$ $\gamma$-ray error boxes surveyed (open
circles) and the known pulsars listed in the public pulsar catalog
(solid dots) \citep{mht+05}. The dashed lines correspond to Galactic
latitudes $\pm 5^{\circ}$, the latitude limits of the Parkes Multibeam
Survey \citep{mlc+01}, which had a comparable sensitivity to the
survey described here.  The centers of the surveyed $EGRET$ targets
lie outside this region.}
\label{fig-1}
\end{figure}

\begin{figure}
\centerline{\psfig{figure=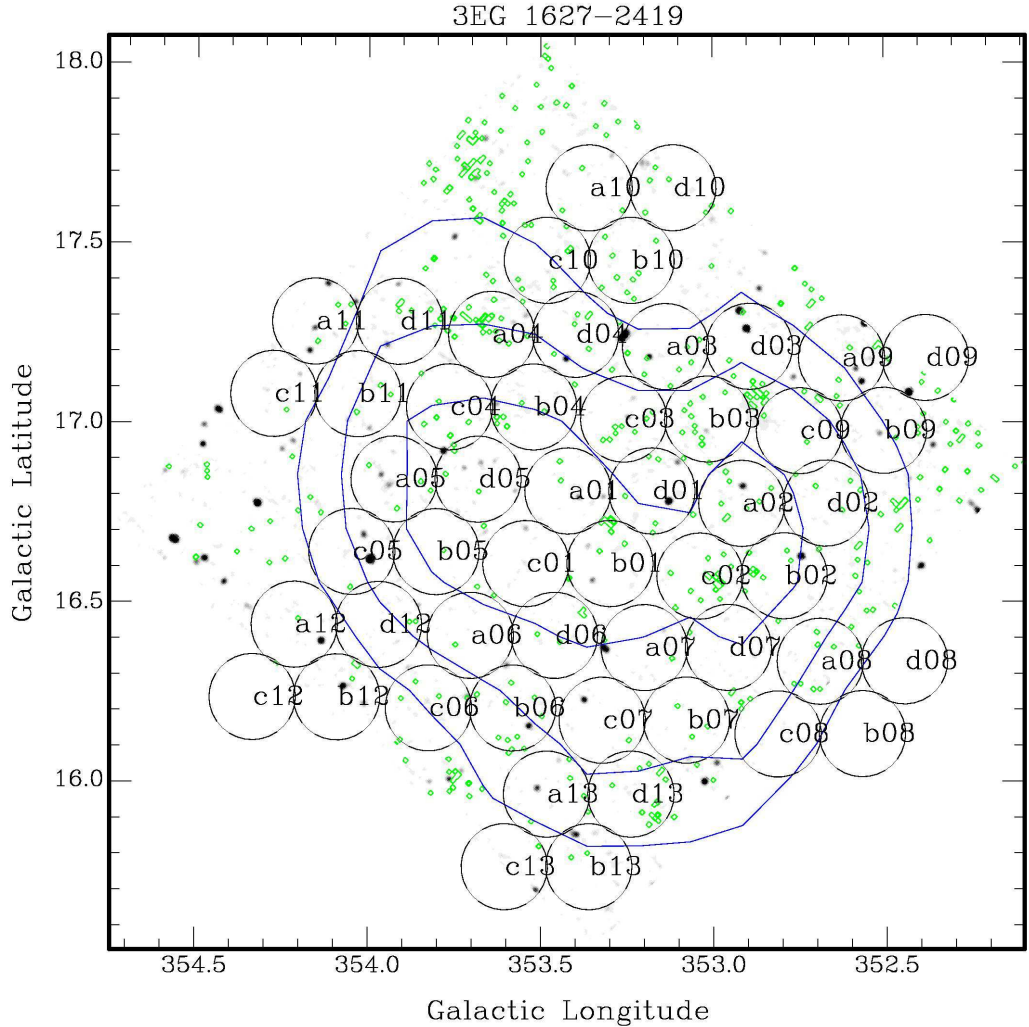,width=6in}}
\caption{Target $EGRET$ source 3EG J1627$-$2419, showing the
$\gamma$-ray error box (contour lines), the multibeam survey coverage
in our search for radio pulsations (circles), X-ray emission from the
$ROSAT$ All-Sky Survey (pixelated squares), and 1.4 GHz emission from
the NRAO VLA Sky Survey (grayscale) \citep{ccg+98}. The radio and
X-ray images were obtained from NASA's {\it SkyView} facility
(http://skyview.gsfc.nasa.gov). The contours represent 68\%, 95\%, and
99\% uncertainties in the $\gamma$-ray source position, and the
circles indicate the Parkes half-power beam size. Four tiled multibeam
pointings are shown (labeled a,b,c,d) with 13 beams each.}
\label{fig-2}
\end{figure}

We used four criteria in the selection of target $EGRET$ sources for
our survey. First, a source was included only if it was not in the
range of the Parkes Multibeam Survey \citep{mlc+01}, which covered
Galactic latitudes $|b| < 5^{\circ}$. Since our targeted survey had a
comparable sensitivity to the Parkes Multibeam Survey, there was no
reason to repeat that coverage.  Second, a source had to have no
strong candidate for an active galactic nucleus (AGN) as determined by
the study of \citet{mhr01}. Third, a source had to have been easily
observable by the Parkes telescope, corresponding to a declination
range $\delta < +20^{\circ}$. Finally, the positional uncertainty from
the 3EG catalog had to be sufficiently small that a single
four-pointing tessellation pattern with the multibeam receiver would
cover virtually the entire 95\% confidence region of the source. Using
these criteria, we selected 56 unidentified $EGRET$ $\gamma$-ray
sources to survey. Figure \ref{fig-1} shows the sky locations of the
56 target sources and the locations of known pulsars. Table
\ref{tbl-1} lists the 56 $EGRET$ sources with their nominal 3EG
positions. These positions were used as the target centers in the
first pointing of each pointing cluster. Since the beams of the
multibeam receiver are spaced two beamwidths apart, four pointings are
required for full coverage of a region on the sky (e.g., Manchester et
al.~2001\nocite{mlc+01}).  This is illustrated in Figure \ref{fig-2}.

We recorded a total of 3016 beams in the survey between June 2002 and
July 2003.\footnote{Nine telescope pointings were repeated in the
survey, and one pointing was missed.  All other pointings were unique
(see Table \ref{tbl-1}).}  For each telescope pointing, we used a
35-minute observation sampled at 0.125 ms with 1-bit per sample.  96
contiguous frequency channels of 3 MHz each were recorded during each
observation, providing a total observing bandwidth of 288 MHz centered
at 1374 MHz. The observing setup was similar to the one described in
detail by \citet{mlc+01} for the Parkes Multibeam Survey, except that
twice the sample rate was used here in order to increase sensitivity
to MSPs. Each resulting beam contained $\sim 200$~MB of raw data,
corresponding to a total of $\sim 600$~GB of raw survey data to be
processed for pulsar signals.

The raw data from the survey were originally processed at McGill
University using the Borg computer cluster and the PRESTO suite of
pulsar analysis tools \citep{r01,
rem02}\footnote{http://www.cv.nrao.edu/$\sim$sransom/presto} with
acceleration searches.  In the search, we dedispersed each data set at
150 trial dispersion measures (DMs) ranging from 0 to 542 pc
cm$^{-3}$, which easily encompassed the expected maximum DM for
Galactic pulsars in the directions observed \citep[][see
Table~1]{cl02}. The values of the DM trials were chosen such that the
spacing did not add to the dispersive smearing already caused by the
finite frequency channels.  Since radio frequency interference (RFI)
can mask pulsar signals, we searched for RFI in particular spectral
channels and time bins for each observation, and a mask was created to
exclude these data from the subsequent reduction and analysis.
Typically about 10-20\% of the data were rejected in this process.

For each trial DM, we summed the frequency channels with appropriate
delays to create a time series. The time series was then Fourier
transformed using a Fast Fourier Transform (FFT), and a red noise
component of the power spectrum (i.e., low-frequency noise in the
data) was removed. This was done by dividing the spectral powers by
the local median of the power spectrum, increasing the number of bins
used in the average logarithmically with frequency.  We masked known
interference signals in the power spectrum, corresponding to less than
0.05\% of the spectrum, and used harmonic summing with up to 8
harmonics to enhance sensitivity to highly non-sinusoidal signals.  In
the acceleration search, we were sensitive to signals in which the
fundamental drifted linearly by up to 100 Fourier bins during the
course of the observation, providing sensitivity to pulsars in tight
binaries; the maximum detectable acceleration was $a_{\rm max} = 6.8
P$ m s$^{-2}$, where $P$ is the pulsar spin period in
milliseconds. This is about 40\% of the maximum acceleration searched
in the Parkes Multibeam Survey processing, which used a segmented
linear acceleration search \citep{fsk+04,l05}. We estimate that our
acceleration search would have been sensitive to all but one of the
known pulsars in double neutron star binary systems (the one exception
being PSR J0737$-$3039A).  We performed folding searches around
candidate periods and period derivatives and examined the results by
eye.  The characteristic signal of interest was a dispersed, wideband,
extremely regular series of pulsations.

\begin{figure}
\centerline{\psfig{figure=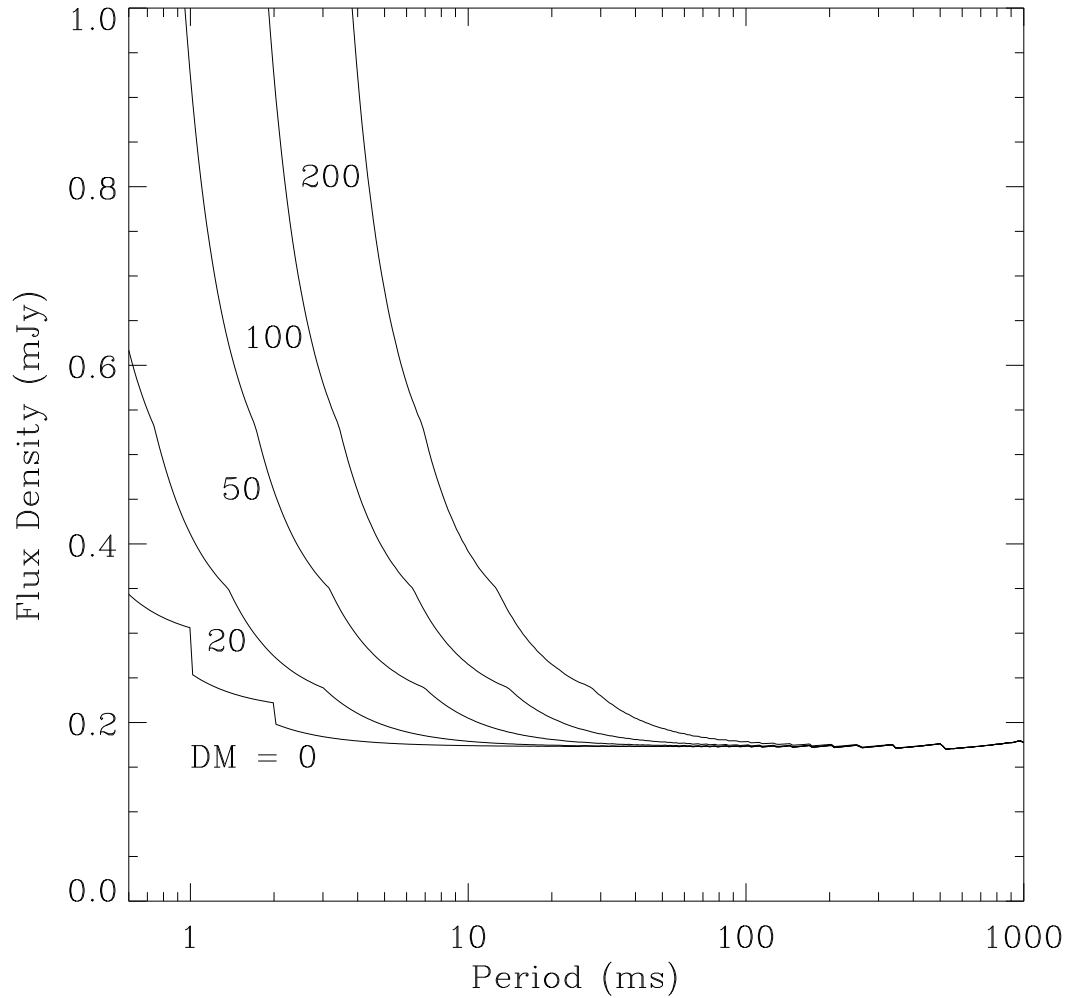,width=6in}}
\caption{Minimum detectable 1400-MHz flux density (in the absence of
RFI) as a function of pulsar period for our survey of $EGRET$
targets. A range of DMs was assumed in the calculation, with the
sensitivity curve for each DM labeled (in units of pc cm$^{-3}$). An
intrinsic duty cycle of 5\% for the pulsed emission was assumed in the
sensitivity calculation as was a sky temperature of 5~K at 1400 MHz;
this is the maximum sky temperature for any of our sources
\citep{hss+82}. In the calculation, we used the gain of the center
beam of the multibeam receiver, which is the most sensitive of the 13
beams.  Averaging over the gains of the 13 beams of the receiver
slightly increases the baseline limit to $\sim 0.2$~mJy. Assuming a
duty cycle smaller than 5\% lowers it.  The inclusion of higher-order
harmonics in the search is the cause of the sudden jumps in the
sensitivity curves at small periods.  The details of the observing
system parameters and the sensitivity calculation, which is for a
blind FFT search, are outlined in \citet{c00} and \citet{mlc+01}. For
the second processing run using the resampled data, the baseline limit
of $\sim 0.2$~mJy remains, but the sensitivity to pulsars with periods
below about 20 ms is sharply degraded for all DMs (see Section
2). Note that a significant red noise component in the FFT from RFI
begins to degrade the sensitivity for periods $\ga 200$~ms and is not
included in the model of the sensitivity.}
\label{fig-3}
\end{figure}

Averaged over the survey, the sensitivity to pulsars in an RFI free
environment was $\sim 0.2$ mJy for most periods and DMs (see Figure
\ref{fig-3}). The sensitivity calculation is outlined in \citet{c00}
and \citet{mlc+01} and was determined for a blind FFT search. RFI
tends to introduce sporadic, highly variable red noise in the power
spectra, especially at low dispersion measures (DM $\la 10$ pc
cm$^{-3}$).  Therefore, sensitivity to slow pulsars ($P \ga 200$~ms)
with low DMs is reduced in a way which is difficult to quantify.  In
addition, the DM peaks of long-period pulsars are broader than those
of MSPs and hence are more difficult to distinguish from zero DM when
the DM is very low.  During this first processing run, we discovered
six new pulsars and redetected all previously known pulsars that were
within the full-width half-maximum area of the survey beams (see Table
\ref{tbl-2}).

We conducted a second processing pass at Haverford College using the
pulsar search packages SEEK and SIGPROC (e.g., Lorimer et al
2000\nocite{lkm+00}).\footnote{http://sigproc.sourceforge.net} The
re-processing of the data with a different analysis package aimed to
see whether there were pulsars that were missed during the first
processing pass. Of particular interest were long-period pulsars ($P
\ga 20$ ms), since fewer than expected were found in the first
processing run.  We therefore decimated the data prior to processing
to reduce their size and thus significantly decrease the processing
time while still maintaining sensitivity to longer-period pulsars.
The data were decimated by a factor of four in frequency and a factor
of 16 in time, resulting in effective frequency channels of 12~MHz
sampled every 2.0~ms.  This reduced the size of each data set by a
factor of 64.  We were in practice sensitive to pulsars with periods
greater than about 20 ms in the re-processing of the data.

These data were dedispersed at 450 trial DMs between 0 and 700 pc
cm$^{-3}$. The large number of DM trials ensured that no weak
candidates with fast periods ($P \sim 20$-30 ms) were missed between
DM steps.  Each resulting time series was Fourier transformed, excised
of RFI, and searched for candidate signals. We then dedispersed and
folded the raw data at DMs and periods around the candidate values. We
redetected all of the pulsars that had been detected in the first
processing run (except for PSR J1614$-$2230, which has a period of
$\sim 3$~ms), but no additional pulsars were found. We also searched
the data for dispersed single pulses. Dispersed radio bursts have
recently been observed from a newly discovered class of transient
radio sources; these sources are believed to be associated with
rotating neutron stars \citep{mll+06}. Our single pulse search
revealed no new candidates, but several known pulsars were redetected
in this way.  We also constructed an archive of the raw data from the
survey on DVD \citep{ccd+04}. A complete index of the survey and
instructions for requesting raw data from the archive is accessible
via the world wide web.\footnote{http://cs.haverford.edu/pulsar}

\section{Results}

We detected a total of 13 pulsars in the survey, six of which were
new. Timing observations quickly established that three of the six new
pulsars are isolated and three are in binary systems.
Table~\ref{tbl-2} lists all 13 pulsars detected in the survey.

\begin{figure}
\centerline{\psfig{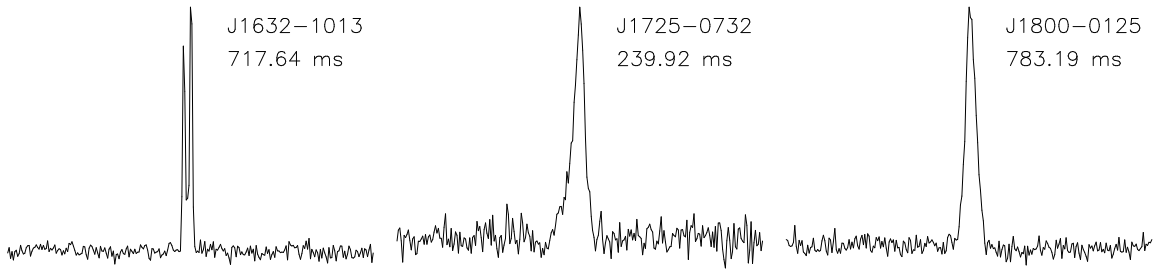}}
\caption{Integrated 20-cm profiles for PSRs J1632$-$1013,
J1725$-$0732, and J1800$-$0125, the three isolated pulsars discovered
in the survey. Each profile is the sum of many timing observations and
has a total of 256 bins. One full period is shown in each case. Timing
parameters for these pulsars, including flux densities and pulse
widths, are presented in Table \ref{tbl-3}.}
\label{fig-4}
\end{figure}

The three new isolated pulsars, PSRs J1632$-$1032, J1725$-$0732, and
J1800$-$0125, were timed at Parkes in 2003 and 2004 with some
supplemental observations taken with the Green Bank Telescope
(GBT). We conducted timing observations at roughly monthly intervals
at several central observing frequencies (mostly 1374 MHz, but also
680, 820, 1400, 1518, and 2934 MHz, depending on the receivers
available at different times) and produced times-of-arrival from the
observations.  The observing setup was similar to the one used for
timing pulsars discovered in the Parkes Multibeam Survey
\citep{mlc+01}.  These data were fit to a model which included spin
parameters, sky position, and DM using the TEMPO software
package.\footnote{http://www.atnf.csiro.au/research/pulsar/tempo} We
used supplemental GBT observations taken in the middle of 2004 along
with the original Parkes survey observations to obtain phase-connected
timing solutions which spanned more than a year.  Table~\ref{tbl-3}
gives the full timing solutions for these three new isolated pulsars
(including 1400-MHz flux densities), and Figure~\ref{fig-4} shows
their 20-cm pulse profiles.

The three new binary pulsars, PSRs J1614$-$2315, J1614$-$2230, and
J1744$-$3922,\footnote{One of the new binary pulsars, PSR
J1744$-$3922, was independently discovered in the re-processing of the
Parkes Multibeam Survey data \citep{fsk+04}.} were regularly timed
with Parkes and the GBT over a similar period of time
\citep{hrr+05}. These pulsars will be discussed in detail by
\citet{rrh+06}. We also detected a fourth binary pulsar, PSR
J0407+1607, in the survey. This pulsar was previously discovered in an
Arecibo drift scan survey by \citet{lxk+05}.

If the pulsar distances estimated from the DMs using the NE2001
Galactic electron density model \citep{cl02} are approximately correct
(to within about a factor of two), then none of the pulsars detected
has a spin-down luminosity which is large enough to clearly account
for the $\gamma$-ray luminosity of its coincident $EGRET$ source. Only
the MSP PSR J1614$-$2230 has a spin-down luminosity of a similar
magnitude to the estimated $\gamma$-ray luminosities of our sources,
which, given the DM distances and $EGRET$ fluxes, are in the $10^{34}$
to $10^{35}$ erg s$^{-1}$ range.  Even PSR J1614$-$2230 would have to
be highly efficient to be the counterpart to its coincident
$\gamma$-ray source (this will be discussed in more detail by Ransom
et al. 2006\nocite{rrh+06}). Therefore, none of the pulsars is a
strong candidate for an $EGRET$ association based on its spin-down
luminosity.  All of the DM-estimated distances to the detected pulsars
($d \ga 1.3$~kpc; see Table \ref{tbl-2}) are too large to be part of a
Gould Belt population, which is expected to have a distance $\la
0.5$~kpc.  In fact, one of the new pulsars, PSR J1632$-$1013, has a DM
which is larger than the maximum expected DM along its line of sight.
Although only about half of the surveyed $EGRET$ sources were within
$30^{\circ}$ of the Galactic center, only PSR J1821+1715 and the
long-period binary PSR J0407+1607 were detected outside this region.

\section{Discussion}

The majority of identified $EGRET$ sources at high Galactic latitudes
are of the blazar sub-class of AGN. As stated above, we selected
against these sources based on the work of \citet{mhr01}. However,
more recent radio and optical work by Sowards-Emmerd and collaborators
\citep{srm03,srm+04} on the complete sample of 3EG sources north of
$-40^{\circ}$ declination has significantly expanded the number of
potential AGN identifications. 33 sources remaining with no potential
AGN counterparts (corresponding to roughly half of all such
unidentified sources at Galactic latitudes $|b| > 5^{\circ}$) were
included in our search. We included about one quarter of the sources
with only weak AGN candidates by their criterion as well. Six of our
sources were identified in their work as having firm AGN associations
(see Table \ref{tbl-1}).  Therefore, for discussion purposes, we
assume that 50\% of all unidentified Galactic sources with $|b| >
5^{\circ}$ were covered in our survey.

One well-discussed model suggests that the mid-latitude $EGRET$
sources are primarily nearby, middle-aged pulsars born in the Gould
belt. This has been motivated by an apparently statistically
significant spatial correlation between the unidentified $\gamma$-ray
sources and the Gould belt \citep{gmb+00,gre01}. \citet{gvh04} have
modeled the pulsar population using estimated pulsar birth rates in
the Gould Belt in addition to simulating the Galactic population as a
whole, and their simulations suggest that $\sim 15$ pulsars ought to
be detectable by $EGRET$ at mid-latitudes, roughly half of which are
radio loud (assuming a particular luminosity law and beaming model for
the radio emission which is consistent with the total known population
of isolated radio pulsars). However, since their simulation accounts
for only $\sim 1/4$ of the total unidentified $\gamma$-ray population,
the hypothesis that all of the sources are pulsars would suggest that
$\sim 15$ radio loud pulsars ought to have been detectable in our
sample of $EGRET$ sources.  A similar study by \citet{czlj04}, based
on the outer gap emission model, finds $\sim 4$ radio loud pulsars
from the Gould belt and another 4 from the remainder of the Galaxy at
$|b| > 5^{\circ}$. The total number of pulsars at mid-latitudes from
this simulation accounts for $\sim 1/2$ the total unidentified
population, indicating that our survey should have detected $\sim 8$
associated radio pulsars. Both of these simulations were done using
estimates of the limiting sensitivities of a variety of previous radio
surveys which were mostly performed at $\sim 400$ MHz and do not
include the various multibeam surveys at mid- and high-latitudes.  Our
survey covered $\sim 50$\% of the potential $EGRET$ pulsars at $|b| >
5^{\circ}$, and yet no plausible radio candidates were discovered.
The absence of detections in our survey is significant given the
discrepancy between our results and the $\sim 8$ and $\sim 15$
detectable radio pulsars predicted in the two models under the
assumption of a single source class consisting of pulsars.  For a
source distance of 0.5 kpc, our 1400-MHz luminosity limit was about
0.05~mJy~kpc$^{2}$; the radio luminosity, $L_{1400}$, is defined as
$L_{1400} = S_{1400} d^{2}$, where $S_{1400}$ is the 1400-MHz flux
density and $d$ is the pulsar distance.  This luminosity limit is
lower than the 1400-MHz luminosity of all but two pulsars for which
this quantity has been measured and published
\citep{mht+05}.\footnote{http://www.atnf.csiro.au/research/pulsar/psrcat}
The surveys used for the studies mentioned above were typically $\sim
4$ times less sensitive than our survey (assuming an average spectral
index of $-2$ for pulsars, as was assumed by Cheng et
al. 2004\nocite{czlj04}). Our results suggest that the simulations
significantly overestimate the radio-loud $\gamma$-ray pulsar
population at mid-latitudes and do not support the hypothesis that
middle-aged, nearby pulsars make up the majority of the unidentified
sources.

There are several important caveats to this conclusion. The first is
that the average radio spectral index of middle-aged, $\gamma$-ray
emitting pulsars is unknown. If, for whatever reason, these sources
preferentially have very steep radio spectra, we might not be
sensitive to them at the relatively high observing frequency of this
survey. The second caveat is the difficulty in distinguishing a peak
at a small but nonzero DM in the data at this frequency.  A clear
indication of a dispersed signal is one of the important ways of
distinguishing a celestial signal from local RFI. Since Gould belt
pulsars are expected to be very close to Earth ($d \la 0.5$ kpc), the
expected DM is less than about 10 pc cm$^{-3}$ along many lines of
sight. This often cannot be clearly differentiated from zero DM with
the high observing frequency of the multibeam system.  This is
especially true of long-period pulsars. In fact, we detected a large
number of promising candidates with pulsar-like characteristics which
peaked at a DM of zero. Although we attempted (and failed) to confirm
some of the most pulsar-like of these candidates at 680 MHz, we still
cannot definitely rule out that some of these candidates may be
astronomical sources.  Observations of these sources at lower
frequencies (300-400 MHz) with modern, wide-bandwidth systems (50-64
MHz) may be able to resolve these low-DM and spectral index
issues. However, a recent 327-MHz search of 19 mid-latitude $EGRET$
error boxes visible from the Arecibo telescope found no new pulsar
counterparts \citep{cml05}, lending support to the conclusion that
pulsars are not powering the majority of these $\gamma$-ray sources.

Although this survey detected more pulsars in binary systems per
square degree (0.032 deg$^{-2}$) outside of globular clusters than any
previous survey, PSR J1614$-$2230 was the only MSP we detected which
is even a marginal counterpart candidate. Recent modeling of
high-energy spectra of MSPs \citep{hum05} suggests that most MSPs
visible to $EGRET$ would be active radio pulsars with significant
radio luminosity. Therefore, the number of observable radio MSPs
detectable by our survey should only depend on the relative radio and
$\gamma$-ray beaming fractions. At large DMs (DM $\ga 100$ pc
cm$^{-3}$), our sensitivity to MSPs is severely compromised owing to
dispersive smearing. However, Table \ref{tbl-1} indicates that less
than half of our $EGRET$ targets have a maximum expected DM greater
than 100 pc cm$^{-3}$, and, of these, only the most distant pulsars
near the edge of the Galactic electron layer would actually have such
large DMs. Dispersive smearing is therefore likely not the reason why
a majority of MSPs would have been missed in our survey. For a
distance of $\sim 3$~kpc, most of the $\gamma$-ray sources would have
luminosities of $\sim 10^{35}$ ergs s$^{-1}$, and so we deem it
unlikely that MSPs could be powering $EGRET$ sources at distances much
further than this. At 3~kpc, our 1400-MHz luminosity limit for a 2-ms
pulsar with a DM of 50 pc ${\rm cm}^{-3}$ is $\sim 5$~mJy
kpc$^2$. While the dependence of radio luminosity on spin-down
luminosity is not well known for MSPs, this level of sensitivity would
have allowed us to detect the majority of known MSPs.  Therefore, our
results do not support the hypothesis that recycled pulsars having
radio luminosities similar to those of the known population make up
the majority of the unidentified $EGRET$ source population.  On the
other hand, the detection of a total of four binary systems in this
survey indicates that deeper surveys for binary pulsars, especially
within $30^{\circ}$ of the Galactic center, appear warranted.

The detection of only 3 new isolated pulsars was somewhat surprising,
especially since we discovered an equal number of new binary pulsars
and detected 6 previously known isolated pulsars within the survey
area (Table \ref{tbl-2}). Since our survey was $\sim 3$ to 4 times
more sensitive than previous surveys (assuming a typical spectral
index), we might have expected to discover a dozen or so new isolated
pulsars.  As noted above, most of the previous surveys at high
latitudes were conducted at lower observing frequencies, and therefore
such a simple estimate is subject to uncertainties in the spectral
index and the influence of RFI. However, the strong detections of all
previously known pulsars argues that these uncertainties may not be
very significant.

We therefore estimate the total number of pulsars we could expect to
detect at our observing frequency by comparing our results with those
of the Swinburne mid-latitude surveys \citep{ebs+01,jac04}. These
surveys covered Galactic longitudes $-100^{\circ} < l < 50^{\circ}$
using the Parkes multibeam receiver and an identical observing setup
to ours, but with only 1/8 the integration time.  The first of these
surveys covered Galactic latitudes $5^{\circ} < |b| < 15^{\circ}$ and
detected 170 pulsars, including 12 binaries. By simply scaling by the
area covered in this survey, the integration time, and assuming a
${\rm d}\log N / {\rm d}\log S$ distribution of $-1$ for Galactic
plane pulsars at 20~cm \citep{bam+03}, we would expect to have
detected a total of $\sim 24$ pulsars instead of 13.  However, we
should have detected only 2-3 binary pulsars, while we detected 4.
The second Swinburne survey, covering $15^{\circ} < |b| < 30^{\circ}$,
detected only 62 pulsars, 11 of which were binaries
\citep{jac04}. This, along with the fact that 11 of our 13 detections
were within $\sim 30^{\circ}$ of the Galactic center, suggests a
strong spatial dependence to the pulsar population out of the plane,
which is hardly surprising. We therefore calculated the number of
isolated pulsars we would have expected to detect within the error
boxes overlapping the coverage of the Swinburne surveys given the
total area covered by our survey within each Swinburne survey and
within $|l| < 30^{\circ}$. Scaling from the surveys and assuming a
${\rm d}\log N / {\rm d}\log S$ distribution of $-1$, we should have
detected $\sim 7$ isolated pulsars but only $\sim 1$ binary pulsar,
when we actually detected 8 and 3, respectively, in this region. In
the $EGRET$ boxes within the Swinburne latitudes but outside their
longitude range (presuming no further longitudinal dependence for
$|l|> 30^{\circ}$), we would have expected $\sim 1$ isolated and $\sim
0$ binary pulsars, while we detected 1 of each.  At higher latitudes,
if the detection rate remained the same for $|b|> 30^{\circ}$ as for
the second Swinburne survey ($15^{\circ} < |b| < 30^{\circ}$), we
would have expected to detect $\sim 2$ pulsars. No pulsars were
detected in our survey at high latitudes.  We therefore conclude that
our results are consistent with an extrapolation from the Swinburne
observations only if we take into account a strong latitudinal
dependence of the isolated pulsar distribution, as expected for a disk
based population, and the apparent concentration of binary pulsars
within $\sim 30^{\circ}$ of the Galactic center. This supports the
trend in the spatial distribution of MSPs suggested by \citet{bjd+06}
obtained by combining data from the Parkes High-Latitude pulsar survey
and the two Swinburne surveys.  This suggests that we have not yet
reached the lower luminosity limit of either the isolated or binary
pulsar populations at mid-Galactic latitudes toward the Galactic
center, since we found approximately what would be expected from a
simple ${\rm d}\log N / {\rm d}\log S$ extrapolation. However, we may
be reaching the luminosity limit toward the anti-center.

\section{Conclusions}

There are now 20 pulsars that are known to lie within 1.5 times the
radius of the 95\% confidence contours of $EGRET$ sources at $|b| >
5^{\circ}$. Of these, only the Crab pulsar and PSR B1055$-$52 have
confirmed associations with the coincident $\gamma$-ray emission. Of
the remaining 18 pulsars, including the 13 detected in our survey and
the recently discovered PSR J2243+1518 \citep{cml05}, none is
energetic enough for a clear association. Other than PSR J1614$-$2230,
which is at best a marginal candidate, no pulsars from any survey have
been found which can be associated with unidentified $EGRET$ error
boxes at mid-Galactic latitudes.  Non-pulsar source classes should
therefore be investigated further. \citet{gkr05} discuss the viability
of low-mass microquasars as $EGRET$ sources. Recently, there has been
the suggestion that much of the $\gamma$-ray emission at mid-latitudes
is due to gas not being included in the models used for calculating
the $\gamma$-ray background maps \citep{gct05}. In this case, many of
the cataloged sources may not be truly point-like. Regardless, as
suggested by spectral and variability studies of the population (e.g.,
Grenier 2003\nocite{gre03}), the likelihood of pulsars being able to
account for a majority of the cataloged unidentified $EGRET$ sources
at intermediate Galactic latitudes seems remote.

\acknowledgements

This work was supported by the Canada Foundation for Innovation, the
Haverford College Faculty Research Fund, the Haverford Faculty Support
Fund, the Keck Northeast Astronomy Consortium, and the NRAO Foreign
Telescope Travel Grants program.  The Parkes radio telescope is part
of the Australia Telescope, which is funded by the Commonwealth of
Australia as a National Facility operated by CSIRO. VMK is a Canada
Research Chair, and JWTH is an NSERC PGS-D Fellow.  VMK received
support from an NSERC Discovery Grant and Steacie Fellowship
Supplement, and by the FQRNT and CIAR. We have made use of the $ROSAT$
data archive of the Max-Planck-Institut f\"{u}r extraterrestrische
Physik (MPE) at Garching, Germany.  We thank the referee, Fernando
Camilo, for helpful suggestions for the revised manuscript, and Dunc
Lorimer for providing key components of the software used in the
re-analysis (SEEK and SIGPROC). We also thank Andrew Cantino, Allison
Curtis, Saurav Dhital, Steve Gilhool, Megan Roscioli, Gabe Roxby, Ryan
Sajac, Reid Sherman, and Aude Wilhelm for contributions to the data
processing and analysis.

\clearpage

\clearpage

\clearpage

\clearpage

\clearpage 

\begin{deluxetable}{lcccccc}
\footnotesize
\tablecaption{$EGRET$ Sources Surveyed. \label{tbl-1}}
\tablewidth{0pt}
\tablehead{
\colhead{Source} &
\colhead{95\%} &
\colhead{Right}& 
\colhead{Declination, $\delta$} &
\colhead{Galactic} &
\colhead{Galactic} &
\colhead{Maximum} \\

\colhead{Name} &
\colhead{error} &
\colhead{ascension, $\alpha$}& 
\colhead{(J2000)} &
\colhead{latitude, $l$} &
\colhead{longitude, $b$} &
\colhead{expected} \\ 

\colhead{(3EG)} &
\colhead{radius\tablenotemark{a}} &
\colhead{(J2000)}& 
\colhead{(dd:mm:ss)} &
\colhead{(deg.)} &
\colhead{(deg.)} &
\colhead{DM\tablenotemark{b}} \\

\colhead{} &
\colhead{(deg.)} &
\colhead{(hh:mm:ss)}& 
\colhead{} &
\colhead{} &
\colhead{} &
\colhead{(pc cm$^{-3}$)} 
}
\startdata
J0038$-$0949\tablenotemark{c}	& 0.59	& 00:38:57	& $-$09:49:11           & 112.69		& $-$72.44		& 30	        \\
J0159$-$3603	& 0.79	& 01:59:21	& $-$36:03:36		& 248.89		& $-$73.04		& 30		\\
J0245$+$1758\tablenotemark{c}	& 0.66\tablenotemark{e}	& 02:45:26	& $+$17:58:11	        & 157.62		& $-$37.11		& 50		\\
J0348$-$5708	& 0.42\tablenotemark{e}	& 03:48:28	& $-$57:08:23		& 269.35		& $-$46.79		& 40		\\
J0404$+$0700\tablenotemark{c}	& 0.70\tablenotemark{e}	& 04:04:36	& $+$07:00:00	        & 184.00		& $-$32.15		& 50		\\
J0407$+$1710	& 0.71	& 04:07:16	& $+$17:10:48	        & 175.63		& $-$25.06		& 70		\\
J0426$+$1333	& 0.45\tablenotemark{e}	& 04:26:40	& $+$13:33:36 		& 181.98		& $-$23.82		& 70		\\
J0429$+$0337	& 0.55\tablenotemark{e}	& 04:29:40	& $+$03:37:48 	 	& 191.44		& $-$29.08		& 60		\\
J0439$+$1105	& 0.92	& 04:39:14	& $+$11:05:24	       	& 186.14		& $-$22.87		& 70		\\
J0442$-$0033	& 0.65	& 04:42:11	& $-$00:33:00       	& 197.39		& $-$28.68		& 50		\\
J0512$-$6150	& 0.59	& 05:12:36	& $-$61:50:24		& 271.25		& $-$35.28		& 50		\\
J0530$-$3626\tablenotemark{c}	& 0.75	& 05:30:09	& $-$36:26:23		& 240.94		& $-$31.29		& 50		\\
J0556$+$0409	& 0.47	& 05:56:14	& $+$04:09:00        	& 202.81		& $-$10.29		& 120   	\\
J0616$-$3310	& 0.63	& 06:16:36	& $-$33:10:11		& 240.35		& $-$21.24		& 70		\\
J0812$-$0646	& 0.72	& 08:12:33	& $-$06:46:48           & 228.64		& +14.62 		& 90	 	\\
J0903$-$3531	& 0.58	& 09:03:09	& $-$35:31:47		& 259.40		& +7.40  		& 330   	\\
J1134$-$1530	& 0.59	& 11:34:38	& $-$15:30:00		& 277.04		& +43.48 		& 40		\\
J1219$-$1520	& 0.80	& 12:19:16	& $-$15:20:24		& 291.56		& +46.82 		& 40		\\
J1234$-$1318	& 0.76	& 12:34:02	& $-$13:18:36		& 296.43		& +49.34 		& 40		\\
J1235$+$0233	& 0.68\tablenotemark{e}	& 12:35:14	& $+$02:33:35 		& 293.28		& +65.13 		& 30		\\
J1310$-$0517	& 0.78	& 13:10:23	& $-$05:18:00	        & 311.69		& +57.25 		& 30		\\
J1314$-$3431	& 0.56	& 13:14:02	& $-$34:31:12		& 308.21		& +28.12 		& 70		\\
J1316$-$5244	& 0.50\tablenotemark{e}	& 13:16:57	& $-$52:45:00		& 306.85		& +9.93  	  	& 220   	\\
J1457$-$1903	& 0.76	& 14:57:40	& $-$19:03:35		& 339.88		& +34.60 		& 50		\\
J1504$-$1537	& 0.70	& 15:04:47	& $-$15:37:48		& 344.04		& +36.38 		& 50		\\
J1616$-$2221	& 0.53\tablenotemark{e}	& 16:16:07	& $-$22:22:12		& 353.00		& +20.03 		& 100           \\
J1627$-$2419	& 0.65	& 16:27:55	& $-$24:19:47		& 353.36		& +16.71 		& 130   	\\
J1631$-$1018	& 0.72	& 16:31:07	& $-$10:18:00		& 5.55  		& +24.94 		& 80		\\
J1634$-$1434	& 0.49\tablenotemark{e}	& 16:34:07	& $-$14:34:11		& 2.33	        	& +21.78 		& 90	        \\
J1638$-$2749\tablenotemark{d} &	0.62 & 16:38:40	& $-$27:49:47		& 352.25		& +12.59 		& 190	\\
J1646$-$0704	& 0.53\tablenotemark{e}	& 16:46:28	& $-$07:04:47	        & 10.85 		& +23.69 		& 80		\\
J1649$-$1611	& 0.65	& 16:49:40	& $-$16:12:00		& 3.35	        	& +17.80  		& 120   	\\
J1652$-$0223	& 0.73\tablenotemark{e}	& 16:52:04	& $-$02:24:00	        & 15.99 		& +25.05 		& 80		\\
J1717$-$2737	& 0.64	& 17:17:12	& $-$27:37:47		& 357.67		& +5.95  		& 430   	\\
J1719$-$0430	& 0.44	& 17:19:09	& $-$04:30:36	        & 17.80	        	& +18.17 		& 110    	\\
J1720$-$7820	& 0.75	& 17:20:52	& $-$78:20:23		& 314.56		& $-$22.17		& 90		\\
J1726$-$0807	& 0.76	& 17:26:26	& $-$08:07:11	        & 15.52 		& +14.77 		& 150   	\\
J1741$-$2050	& 0.63	& 17:41:38	& $-$20:50:24		& 6.44  		& +5.00  		& 490   	\\
J1744$-$3934	& 0.66	& 17:44:48	& $-$39:34:11		& 350.81		& $-$5.38 		& 470   	\\
J1746$-$1001	& 0.76	& 17:46:00	& $-$10:01:47		& 16.34 	    	& +9.64  		& 250   	\\
J1800$-$0146	& 0.77	& 18:00:52	& $-$01:46:47	        & 25.49 		& +10.39 		& 210   	\\
J1822$+$1641	& 0.77	& 18:22:16	& $+$16:42:00 		& 44.84 		& +13.84 		& 120   	\\
J1825$-$7926	& 0.78	& 18:25:02	& $-$79:26:24		& 314.56		& $-$25.44		& 80            \\
J1828$+$0142\tablenotemark{c}	& 0.55	& 18:28:59	& $+$01:43:12 		& 31.90 		& +5.78  		& 370   	\\
J1834$-$2803	& 0.52	& 18:34:21	& $-$28:03:35		& 5.92  		& $-$8.97 		& 260   	\\
J1836$-$4933	& 0.66	& 18:38:04	& $-$49:33:36		& 345.93		& $-$18.26		& 120   	\\
J1847$-$3219	& 0.80	& 18:47:35	& $-$32:19:11		& 3.21  		& $-$13.37		& 180   	\\
J1858$-$2137	& 0.36\tablenotemark{e}	& 18:58:26	& $-$21:37:12		& 14.21 		& $-$11.15		& 200   	\\
J1904$-$1124	& 0.50	& 19:04:50	& $-$11:24:35		& 24.22 		& $-$8.12 		& 280   	\\
J1940$-$0121	& 0.79	& 19:40:55	& $-$01:21:36           & 37.41 		& $-$11.62		& 170   	\\
J1949$-$3456	& 0.61	& 19:49:09	& $-$34:56:23		& 5.25  		& $-$26.29		& 80		\\
J2034$-$3110\tablenotemark{c}	& 0.73\tablenotemark{e}	& 20:34:55	& $-$31:10:48		& 12.25 		& $-$34.64		& 60		\\
J2219$-$7941	& 0.63\tablenotemark{e}	& 22:19:59	& $-$79:41:24		& 310.64		& $-$35.06		& 50		\\
J2243$+$1509	& 1.04	& 22:43:07	& $+$15:10:12		& 82.69 		& $-$37.49		& 80		\\
J2251$-$1341	& 0.77	& 22:51:11	& $-$13:41:23		& 52.48 		& $-$58.91		& 30		\\
J2255$-$5012	& 0.70\tablenotemark{e}	& 22:55:57	& $-$50:12:35		& 338.75		& $-$58.12		& 40		\\
\enddata

\tablecomments{~Listed positions are the nominal 3EG positions, which
were used as the target centers for the first of four interleaved
pointings for each source.}

\tablenotetext{a}{Values are the radii of circles containing the same
solid angle as the 95\% confidence contours of the sources and were
obtained from the 3EG catalog \citep{hbb+99}.}

\tablenotetext{b}{Estimated from the NE2001 Galactic electron density
model \citep{cl02} and rounded to the nearest tens value.}

\tablenotetext{c}{Identified by \citet{srm03} or \citet{srm+04} as
having a firm AGN association.}

\tablenotetext{d}{One of the four pointings required to cover 3EG
J1638$-$2749 was not observed in the survey.}

\tablenotetext{e}{Obtained by multiplying the 68\% contour
radius by 1.62. This is necessary in cases of unclosed or extremely
irregular 95\% confidence contours \citep{hbb+99}.}

\end{deluxetable} 

\begin{deluxetable}{lrccccl}
\footnotesize
\tablecaption{All Pulsars Detected in the Survey.\label{tbl-2}}
\tablewidth{0pt}
\tablehead{
\colhead{PSR} &
\colhead{$P$} &
\colhead{DM} &
\colhead{Distance\tablenotemark{a}} &
\colhead{$\log \dot{E}$\tablenotemark{b}} &
\colhead{3EG Target} &
\colhead{Notes} \\
\colhead{} &
\colhead{(s)} &
\colhead{(pc cm$^{-3}$)} &
\colhead{(kpc)} & 
\colhead{(erg s$^{-1}$)} &
\colhead{Source} &
\colhead{} 
}
\startdata
J0407$+$1607  &   0.0257  &   36      & 1.3    & 32.26  & J0407$+$1710 & redetected, binary \\
J1614$-$2315  &   0.0335  &   52      & 1.8    & 31.98  & J1616$-$2221 & new, binary \\
J1614$-$2230  &   0.0032  &   35      & 1.3    & 34.09  & J1616$-$2221 & new, binary \\
J1632$-$1013  &   0.7176  &   90      & $> 50$ & 30.85  & J1631$-$1018 & new \\
J1650$-$1654  &   1.7496  &   43      & 1.4    & 31.38  & J1649$-$1611 & redetected \\
J1725$-$0732  &   0.2399  &   59      & 1.9    & 33.09  & J1726$-$0807 & new \\
J1741$-$2019  &   3.9045  &   75      & 1.7    & 31.04  & J1741$-$2050 & redetected \\
B1737$-$39    &   0.5122  &  159      & 3.1    & 32.76  & J1744$-$3934 & redetected \\ 
J1744$-$3922  &   0.1724  &  148      & 3.1    & 31.11  & J1744$-$3934 & new, binary \\
J1800$-$0125  &   0.7832  &   50      & 1.7    & 32.98  & J1800$-$0146 & new \\
J1821$+$1715  &   1.3667  &   60      & 2.8    & 31.11  & J1822$+$1641 & redetected \\
J1832$-$28    &   0.1993  &  127      & 3.5    & 31.80  & J1834$-$2803 & redetected \\ 
J1904$-$1224  &   0.7508  &  118      & 3.3    & 31.84  & J1904$-$1124 & redetected \\ 
\enddata

\tablenotetext{a}{Estimated from the NE2001 Galactic electron density
model of \citet{cl02}.}

\tablenotetext{b}{$\dot{E} \equiv 4\pi^{2}I\dot{P}/P^{3}$, where a
moment of inertia of $I = 10^{45}$~g cm$^{2}$ is assumed.}

\end{deluxetable}

\begin{deluxetable}{lccc}
\footnotesize    
\tablecaption{Timing Parameters for  
Three Newly Discovered Isolated Pulsars.\label{tbl-3}}
\tablewidth{0pt}
\tablehead{
\colhead{Name} & 
\colhead{J1632$-$1013} & 
\colhead{J1725$-$0732} &
\colhead{J1800$-$0125}
} 
\startdata
Right ascension, $\alpha$ (J2000)                                 & $16^{\rm h} 32^{\rm m} 54^{\rm s}.20$(2)         & $17^{\rm h} 25^{\rm m} 12^{\rm s}.281$(6)            & $18^{\rm{h}} 00^{\rm{m}} 22^{\rm s}.08$(3)        \\
Declination, $\delta$ (J2000)                                     & $-10^{\circ} 13^{\prime} 18^{\prime \prime}$(1)  & $-07^{\circ} 32^{\prime} 59^{\prime \prime}.2$(3)    & $-01^{\circ} 25^{\prime} 30^{\prime \prime}.6$(7) \\
Period, $P$ (ms)                                                  & 717.63732795(2)  & 239.919487227(4)       & 783.18548958(3) \\
Period derivative, $\dot{P}$ ($\times 10^{-15}$)                  & 0.066(1)         & 0.4296(3)              & 11.537(5)    \\
Dispersion measure, DM (pc cm$^{-3}$)                             & 89.9(2)          & 58.91(7)               & 50.0(2)      \\
Epoch of period (MJD)                                             & 52820.00         & 52820.58               & 52820.00     \\
RMS residual (ms)                                                 & 2.3              & 0.9                    & 1.6          \\
Number of TOAs                                                    & 91               & 71                     & 65           \\
Timing span (days)                                                & 731              & 587                    & 493          \\
1400-MHz flux density (mJy)\tablenotemark{a}                      & 0.15(5)          & 0.11(3)                & 0.14(4)      \\
FWHM pulse width (\% of $P$)                                      & 2.8              & 4.1                    & 3.5          \\
Characteristic age, $\tau_{c}$ (Myr)\tablenotemark{b}             & 172              & 8.85                   & 1.08         \\
Surface magnetic field, $B$ ($\times 10^{12}$ G)\tablenotemark{c} & 0.220            & 0.325                  & 3.042        \\
Spin-down luminosity, $\dot{E}$ (erg s$^{-1}$)                    & $7.05 \times 10^{30}$          & $1.23 \times 10^{33}$            & $9.48 \times 10^{32}$  \\
\enddata

\tablecomments{Figures in parentheses represent the formal 1$\sigma$
uncertainties (obtained from TEMPO) in the least-significant digit
quoted.}

\tablenotetext{a}{Uncertainties are estimated to be $\sim 30$\% of the
flux value in each case.}

\tablenotetext{b}{$\tau_{c} \equiv P/2\dot{P}$.}

\tablenotetext{c}{$B \equiv 3.2 \times 10^{19} (P\dot{P})^{1/2}$ G.}

\end{deluxetable}


\end{document}